\documentclass[sigconf, natbib=true]{acmart}

\AtBeginDocument{%
  }

\usepackage{graphicx}

\usepackage{amsmath}
\usepackage{caption}
\usepackage{subcaption}
\usepackage{multirow}
\usepackage{booktabs}
\usepackage{tikz}
\usetikzlibrary{shapes,arrows,positioning,calc}
\usepackage{adjustbox}
\usepackage{array}
\usepackage{float}
\usepackage{enumitem}
\usepackage{amssymb}

\usepackage{xcolor}
\usepackage[linesnumbered,ruled,vlined]{algorithm2e}
\DontPrintSemicolon

\SetKwComment{Comment}{\color{green!50!black}// }{}

\SetKwProg{Function}{function}{}{}

\begin{document}

\title{Compress, Cross and Scale: Multi-Level Compression Cross Networks for Efficient Scaling in Recommender Systems}

\author{Heng Yu}
\orcid{0000-0002-2158-862X}
\authornote{Equal Contributions.}
\affiliation{%
  \institution{Bilibili Inc.}
  \city{Shanghai}
  \country{China}
}
\email{yuheng01@bilibili.com}

\author{Xiangjun Zhou}
\authornotemark[1]
\affiliation{%
  \institution{Bilibili Inc.}
  \city{Shanghai}
  \country{China}
}
\email{zhouxiangjun@bilibili.com}

\author{Jie Xia}
\authornotemark[1]
\affiliation{%
  \institution{Bilibili Inc.}
  \city{Shanghai}
  \country{China}
}
\email{xiajie@bilibili.com}

\author{Heng Zhao}
\authornotemark[1]
\affiliation{%
  \institution{Bilibili Inc.}
  \city{Shanghai}
  \country{China}
}
\email{zhaoheng@bilibili.com}

\author{Anxin Wu}
\affiliation{%
  \institution{Bilibili Inc.}
  \city{Shanghai}
  \country{China}
}
\email{wuanxin@bilibili.com}

\author{Yu Zhao}
\affiliation{%
  \institution{Bilibili Inc.}
  \city{Shanghai}
  \country{China}
}
\email{zhaoyu07@bilibili.com}

\author{Dongying Kong}
\authornote{Corresponding Author.}
\affiliation{%
  \institution{Bilibili Inc.}
  \city{Shanghai}
  \country{China}
}
\email{limin07@bilibili.com}

\renewcommand{\shortauthors}{Yu et al.}

\begin{abstract}
Modeling high-order feature interactions efficiently is a central challenge in click-through rate and conversion rate prediction.
Modern industrial recommender systems are predominantly built upon deep learning recommendation models (DLRMs), where the interaction backbone plays a critical role in determining both predictive performance and system efficiency.
However, existing interaction modules often struggle to simultaneously achieve strong interaction capacity, high computational efficiency, and good scalability, resulting in limited ROI when models are scaled under strict production constraints.
In this work, we propose the \textbf{M}ulti-\textbf{L}evel \textbf{C}ompression \textbf{C}ross network (MLCC), a structured feature interaction architecture that organizes feature crosses through hierarchical compression and dynamic composition, which can efficiently capture high-order feature dependencies while maintaining favorable computational complexity.
We further introduce MC-MLCC, a \textbf{M}ulti-\textbf{C}hannel extension that decomposes feature interactions into parallel subspaces, enabling efficient horizontal scaling with improved representation capacity and significantly reduced parameter growth.
Extensive experiments on three public benchmarks and a large-scale industrial dataset show that our proposed models consistently outperform strong DLRM-style baselines by up to \textbf{0.52\%} AUC, while reducing model parameters and FLOPs by up to \textbf{26$\times$} under comparable performance.
Comprehensive scaling analyses demonstrate stable and predictable scaling behavior across embedding dimension, head number, and channel count, with channel-based scaling achieving substantially better efficiency than conventional embedding inflation.
Finally, online A/B testing on a real-world advertising platform validates the practical effectiveness of our approach, which has been widely adopted in Bilibili advertising system under strict latency and resource constraints. Code for this work is publicly available at: \url{https://github.com/shishishu/MLCC}.

\end{abstract}




\maketitle

\section{Introduction}

Click-through rate (CTR) and conversion rate (CVR) prediction lie at the core of modern recommender systems and online advertising platforms.
Their accuracy directly impacts user experience, platform efficiency, and business revenue, making them long-standing and highly active research topics.
In industrial practice, these tasks are predominantly addressed using deep learning recommendation models (DLRMs), which combine large embedding tables with dedicated feature interaction modules to model complex relationships among heterogeneous categorical and numerical features.

Within DLRM-style models, the feature interaction component plays a central role.
It is responsible for transforming sparse and high-dimensional feature representations into informative cross features that downstream networks can effectively exploit.
Over the past decade, a wide range of interaction modeling techniques has been proposed, including inner- and outer-product based operators \cite{qu2016product}, cross networks \cite{wang2017deep,wang2021dcn}, factorization mechanisms, and more recently attention-based or token-level architectures\cite{song2019autoint,zhu2025rankmixer}.
While these methods have achieved notable success, their performance improvements often come at the cost of increased computational overhead, making efficient deployment at scale increasingly challenging.

A key difficulty lies in the inherent trade-offs of existing interaction designs.
Lightweight interaction operators are computationally efficient but lack sufficient capacity to capture complex high-order dependencies.
In contrast, more expressive architectures, such as deep cross networks \cite{wang2021dcn} or attention-based interaction layers \cite{song2019autoint}, provide stronger modeling power but incur rapidly growing costs in parameters, memory footprint, and latency.
Moreover, many interaction modules scale poorly: increasing embedding dimensionality, interaction depth, or feature count frequently leads to high computation growth or unstable optimization.
As a result, it remains challenging for a single interaction architecture to simultaneously achieve high computational efficiency, strong interaction capacity, and favorable scalability.

This limitation becomes particularly pronounced as DLRM-style models continue to scale.
Inspired by advances in large-scale deep learning, practitioners increasingly explore larger embeddings, deeper networks, or more complex interaction modules.
However, without a principled and structured interaction backbone, additional capacity often yields diminishing returns, poor resource efficiency, or unfavorable return-on-investment (ROI) in real-world systems.
These observations raise a fundamental question: \emph{how can we design a feature interaction architecture that is not only expressive, but also inherently efficient and scalable along multiple dimensions?}

In this work, we argue that overcoming this trade-off requires a fundamentally different approach to interaction design.
We propose the \textbf{M}ulti-\textbf{L}evel \textbf{C}ompression \textbf{C}ross network (\textbf{MLCC}), a novel interaction architecture built around a unified paradigm of \emph{Compress, Cross, and Scale}.
Rather than directly operating on high-dimensional embeddings, MLCC  applies hierarchical compression to construct compact and structured representations.
Feature interactions are then modeled through a dynamic and learnable cross mechanism that generalizes traditional similarity-based operators.
This design enables MLCC to capture high-order feature interactions efficiently while maintaining favorable computational complexity.

Building on MLCC, we further introduce its \textbf{M}ulti-\textbf{C}hannel extension \textbf{MC-MLCC}, which unlocks an additional horizontal scaling pathway.
Instead of relying solely on increasing embedding dimensionality, MC-MLCC distributes interaction modeling capacity across multiple parallel channels.
Each channel learns complementary interaction subspaces, allowing the overall model expressiveness to grow with significantly lower parameter and computation overhead.
This channel-based scaling provides a practical and robust alternative when embedding inflation becomes inefficient or difficult to optimize.

We conduct extensive experiments on three public benchmarks and a large-scale industrial dataset to evaluate the effectiveness and scalability of the proposed architectures.
Under matched or lower computational budgets, MLCC consistently outperforms strong DLRM-based baselines, while MC-MLCC further improves performance by exploiting channel-level scaling.
Through comprehensive ablation studies and unified scaling analyses, we demonstrate that the proposed designs exhibit stable and predictable scaling behavior across embedding size, head number, and channel count.
Moreover, online deployment on a real-world e-commerce advertising platform confirms that our approach delivers substantial performance gains without violating strict latency and resource constraints.

In summary, this paper makes the following contributions:
\begin{itemize}
    \item We propose \textbf{MLCC}, a structured feature interaction architecture that combines hierarchical compression with dynamic cross modeling to achieve strong expressiveness and computational efficiency.
    Across three public benchmarks and a large-scale industrial dataset, MLCC consistently outperforms strong DLRM-style baselines by +0.07\% to +0.20\% AUC, while achieving up to 6$\times$ reduction in parameters and computation compared with embedding-scaled or deeply stacked models.
    \item We further introduce \textbf{MC-MLCC}, a multi-channel extension that enables efficient horizontal scaling and provides a new perspective on increasing interaction capacity beyond embedding dimensionality.
    On the industrial dataset, MC-MLCC matches the performance of top baselines while requiring over 26$\times$ fewer parameters and FLOPs, and surpasses computation-matched DLRM variants by up to +0.52\% AUC on public datasets.
    \item We conduct comprehensive scaling analyses along multiple dimensions, including head number, embedding dimension, and channel count.
    Results demonstrate that channel-based scaling yields substantially better ROI than conventional embedding inflation.
    \item Moreover, online A/B testing on a real-world advertising platform validates the practical impact of the proposed framework, delivering a cumulative +32\% ADVV\footnote{Advertiser Value (ADVV), defined as conversions weighted by advertiser bids, is a key online business metric in advertising that measures the value delivered to advertisers.
} improvement under strict latency constraints.
\end{itemize}

Together, these contributions show that carefully designed interaction architectures can serve as a principled foundation for scalable and efficient recommendation models in industrial environments.

\section{Related Work}

\subsection{Feature Interaction Modeling}

Recommender systems have seen extensive research on modeling feature interactions, moving from manual cross-features to automated learning of feature interaction structures.

Early neural approaches like the Product-based Neural Network (PNN) family \cite{qu2016product} introduced explicit product layers (inner or outer products) to directly capture pairwise interactions.
Models such as Wide \& Deep \cite{cheng2016wide} and  Deep Factorization Machine (DeepFM) \cite{guo2017deepfm} combined a ``wide'' component for memorization and a ``deep'' component for generalization, enabling both low-order and high-order feature learning in one network.

 Deep \& Cross Network (DCN) \cite{wang2017deep} took this further by adding a Cross Network layer to explicitly generate bounded-degree cross features, though its fixed cross formula became less effective at web scale.
DCNv2 \cite{wang2021dcn} addressed this by allowing multiple stacked cross layers and using low-rank factorization, making feature crossing more expressive yet efficient in large-scale settings.
In parallel, xDeepFM \cite{lian2018xdeepfm} introduced a Compressed Interaction Network to explicitly construct higher-order feature combinations layer by layer, while still leveraging an implicit DNN component, yielding strong performance without manual feature engineering. 

Attention mechanisms have also been adopted for interaction modeling: AutoInt \cite{song2019autoint}, for example, uses multi-head self-attention \cite{vaswani2017attention} to automatically learn high-order relationships, improving accuracy and interpretability. To approximate the richness of manual crosses, some works like Co-Action Network (CAN) \cite{bian2020can} propose  Co-Action Units to efficiently capture targeted pairwise interactions.

Most recently, driven by scale and efficiency requirements, new paradigms have emerged. On one hand, Wukong \cite{zhang2024wukong} stacks factorization machine blocks to ensure model capacity grows with depth and width, demonstrating a scaling law for recommendation akin to that of large pretrained models. On the other hand, RankMixer \cite{zhu2025rankmixer} fuses Transformer-style parallelism with a sparse expert design to unify feature crossing modules, achieving an order-of-magnitude increase in model size and FLOPs utilization without extra latency. In summary, feature interaction modeling has evolved from explicit cross layers and implicit DNNs to attention-based and combinatorial architectures, and now to ultra-large-scale parallel and expert models. These advances significantly improve recommendation accuracy while balancing model complexity and deployability, laying a foundation for robust feature interaction modeling in modern recommender systems.

\subsection{Scalability of Recommendation Models}

In the pursuit of scaling up recommendation models, one of the most straightforward directions has been to enlarge embedding tables, as embedding layers typically dominate parameter counts in DLRM-style architectures.
However, recent studies suggest that naively increasing embedding dimensionality does not necessarily lead to proportional gains in recommendation performance.
Empirical analyses show that as embedding size grows, learned representations often concentrate in a low-dimensional subspace, a phenomenon commonly referred to as \emph{embedding collapse}~\cite{guo2024embedding,he2025understanding}.
From an optimization and representation perspective, such collapse manifests as low effective rank or highly imbalanced embedding spectra, implying that additional dimensions contribute little useful information despite increased memory and parameter cost~\cite{loveland2025understanding}.

Similar observations have been reported across different collaborative filtering and recommendation settings, where embedding representations exhibit limited intrinsic dimensionality or severe spectral imbalance~\cite{chen2024towards,peng2025balancing,loveland2025understanding}.
These findings collectively indicate that sparse scaling via embedding enlargement alone is insufficient, unless the model architecture can actively transform raw capacity into diverse and meaningful feature interactions.
Recent analyses further connect this phenomenon to broader scaling behaviors in recommendation models, suggesting that effective scaling requires architectural support rather than parameter growth alone~\cite{ardalani2022understanding}.

Against this backdrop, recent scaling strategies have progressively shifted away from pure embedding inflation toward architectural approaches that better translate model capacity into effective representation power.
Broadly speaking, existing efforts can be grouped into three complementary directions.

First, a line of work focuses on \emph{enhancing feature interaction modules} to improve expressiveness and utilization of embedding capacity without excessive embedding enlargement.
Typical approaches include structured cross networks, compressed interaction operators, and attention-based or hybrid explicit--implicit designs~\cite{wang2017deep,wang2021dcn,lian2018xdeepfm,song2019autoint,bian2020can,zhang2024wukong,zhu2025rankmixer}, which are reviewed in detail in the previous subsection.
Recent analyses further point out that poorly structured or overly deep interaction layers may suffer from \emph{interaction collapse}, motivating more principled interaction architectures~\cite{xu2025addressing}.

Second, another stream of research explores \emph{deepening network architectures} as an alternative scaling axis.
Early DLRM-style models rely on deeper MLP backbones to implicitly capture higher-order feature combinations, while subsequent works show that carefully structured depth scaling can yield more predictable gains.
For example, stacking multiple interaction or factorization-based blocks enables hierarchical abstraction of cross features and improves capacity utilization, revealing stable scaling behavior with depth and width~\cite{zhang2024wukong,ardalani2022understanding}.
Complementary efforts also enhance the scalability of deep interaction models through optimized attention mechanisms and system-level improvements, enabling deeper architectures under practical latency constraints~\cite{xu2024enhancing,ye2025fuxi}.

Third, inspired by the success of large pre-trained models, recent studies reformulate recommendation as a large-scale or generative modeling problem, scaling overall model capacity rather than individual components.
A notable milestone is HSTU~\cite{zhai2024actions}, which introduces a Transformer-based sequential recommender with over one trillion parameters and demonstrates power-law scaling with compute, providing early evidence of LLM-style scaling laws in recommendation~\cite{zhang2024scaling}.
Subsequent systems incorporate domain-specific inductive biases or efficiency techniques—such as reintroducing handcrafted cross features, hierarchical compression, or unified retrieval-and-ranking objectives—to balance scale and practicality~\cite{han2025mtgr,chen2024hllm,rajput2023recommender,deng2025onerec}.

In summary, scaling recommendation models has evolved from sparse embedding expansion to richer architectural scaling along interaction, depth, and overall model capacity dimensions.
While large generative recommenders demonstrate impressive scaling effects, their cost and complexity motivate continued exploration of efficient interaction architectures that can achieve favorable scaling behavior under realistic production constraints.

\section{Methodology}

\subsection{Overview}
\label{subsec:overview}

Traditional cross-feature operations, which explicitly enumerate interactions between fields, incur substantial computational overhead that grows rapidly with feature dimensionality, and often become semantically redundant at scale.
To move beyond such flat and exhaustive schemes, we propose the \textbf{M}ulti-\textbf{L}evel \textbf{C}ompression \textbf{C}ross network (MLCC) framework that progressively abstracts, interweaves, and purifies token representations across multiple semantic levels, following a unified \emph{Compress--Cross--Scale} paradigm.
MLCC can be interpreted as a hierarchical distillation and interaction framework for structured feature representation learning.

The process begins with a set of \emph{original tokens}, directly derived from the raw input feature space.
These tokens are first transformed by a \emph{global compressor} (GC), which produces a condensed set of \emph{global tokens}.
Unlike naive down-sampling, these tokens serve as globally contextualized abstractions that integrate holistic information from the entire token set.

Subsequently, the original tokens and global tokens are fed into a dedicated feature-interaction module, termed \emph{progressive layered crossing} (PLC), which performs structured fusion to generate \emph{interwoven tokens} via dynamic weights.
This stage can be viewed as a conditional transformation that entangles raw token representations with globally compressed context through progressive interactions.
Rather than simply concatenating different levels of representation, the PLC module enables a dynamic exchange between local token-level semantics and global contextual representations, constructing tokens that are semantically richer and structurally more aligned with high-level dependencies.

While this interleaving boosts representational power, it also risks introducing informational redundancy.
To mitigate this, we further employ a \emph{local compressor} (LC), which refines the interwoven tokens through token-wise projection, ultimately producing the final set of \emph{refined tokens}.
This final stage acts as a token-wise distillation mechanism, selectively preserving salient semantics while discarding noise and spurious correlations.

\begin{figure*}
    \centering

  \begin{minipage}[b]{0.53\linewidth}
    \centering
    \includegraphics[width=0.85\linewidth]{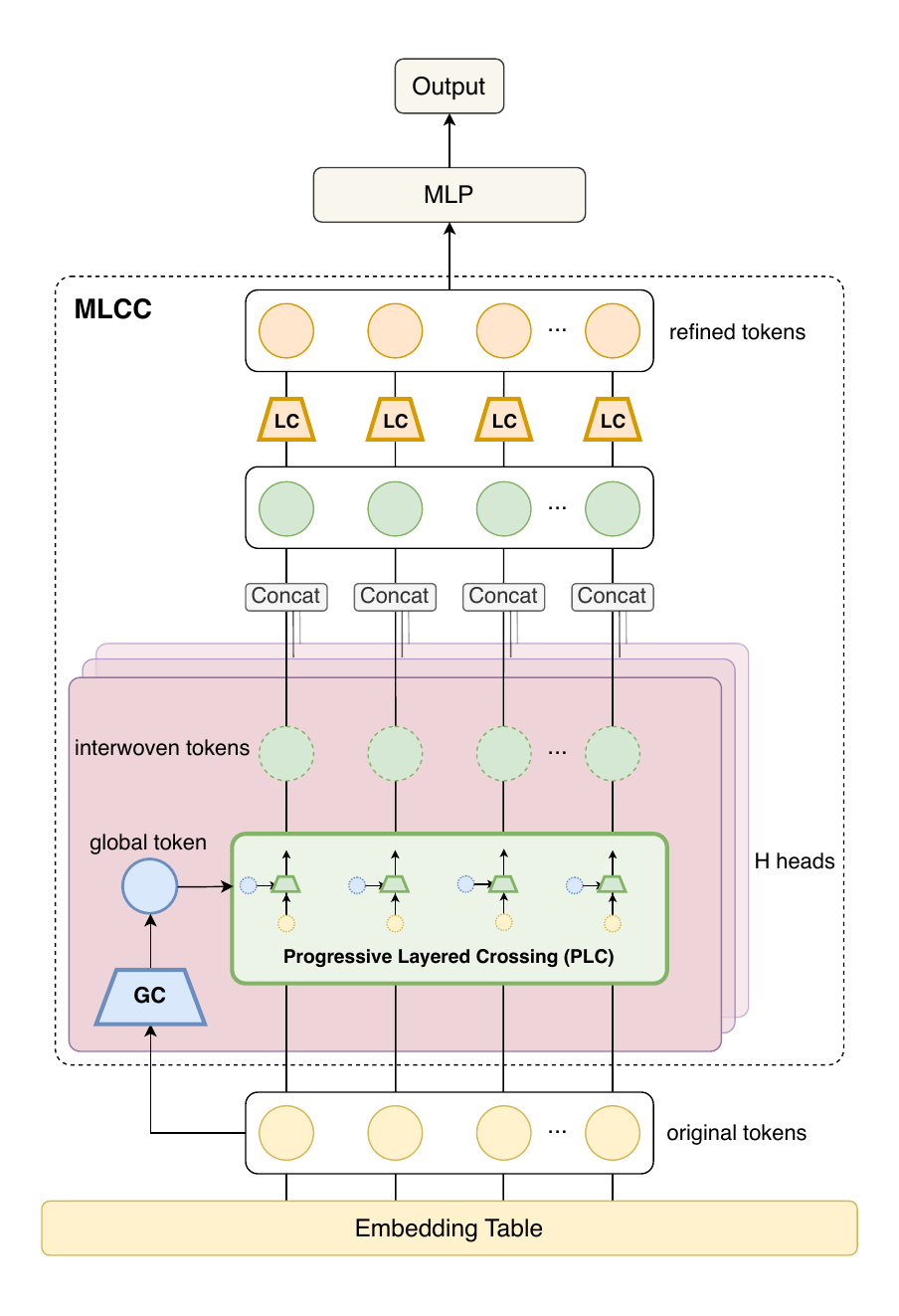}
    \subcaption{MLCC}
    \label{fig:mlcc_main}
  \end{minipage}
  \begin{minipage}[b]{0.43\linewidth}
    \centering
    \begin{minipage}[b]{\linewidth}
        \centering
      \includegraphics[width=0.9\linewidth]{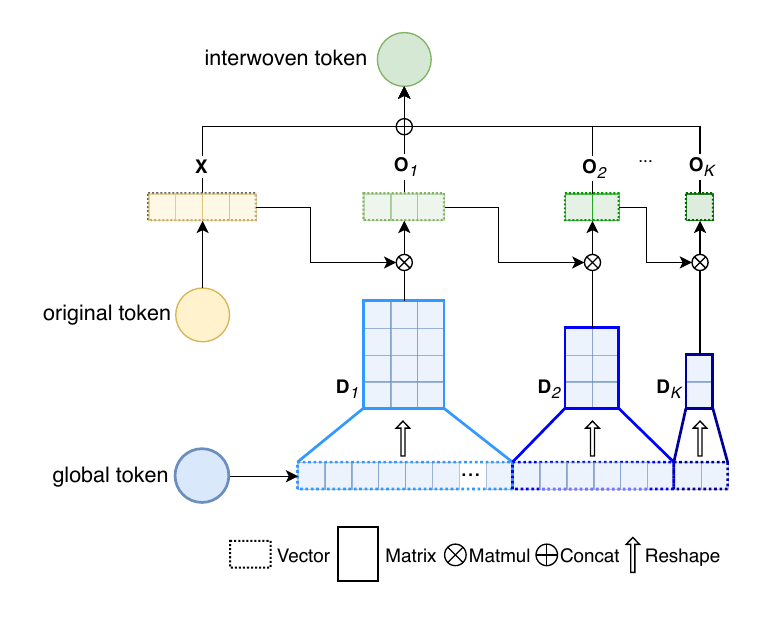}
      \subcaption{PLC}
      \label{fig:mlcc_plc}
    \end{minipage}
    \begin{minipage}[b]{\linewidth}
        \centering
      \includegraphics[width=0.9\linewidth]{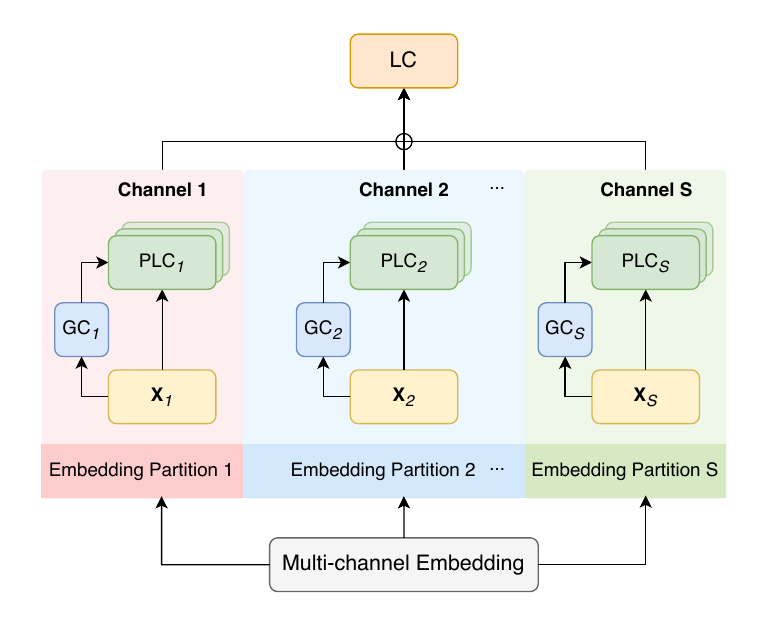}
      \subcaption{MC-MLCC}
      \label{fig:mlcc_multi}
    \end{minipage}
  \end{minipage}
  \captionsetup{skip=3pt}
\caption{Overview of the proposed Multi-Level Compression Cross (MLCC) framework following the compress--cross--scale paradigm. 
(\subref{fig:mlcc_main}) Overall MLCC architecture, consisting of a global compressor (GC), a progressive layered crossing module (PLC), and a local compressor (LC). 
(\subref{fig:mlcc_plc}) Detailed structure of the PLC module, which parameterizes a dynamic MLP using global tokens to model high-order token interactions. 
(\subref{fig:mlcc_multi}) Multi-channel extension of MLCC (MC-MLCC), where multiple GC--PLC pipelines operate in parallel channels and their outputs are aggregated by a shared LC module.}

    \label{fig:mlcc}
\end{figure*}

Formally, the MLCC pipeline can be expressed as:
\begin{align}
\mathbf{M} & :=\text{GC}(\mathbf{X})\label{equ:mlcc_1}\\ 
\mathbf{C} & :=\text{PLC}(\mathbf{X};\mathbf{M})\label{equ:mlcc_2}\\
 \mathbf{X}' & := \text{LC}(\mathbf{C})\label{equ:mlcc_3}
\end{align}
where $\mathbf{X}$, $\mathbf{M}$, $\mathbf{C}$, and $\mathbf{X}'$ denote the original tokens, global tokens, interwoven tokens, and refined tokens, respectively.
An overview of the proposed MLCC framework is illustrated in Figure~\ref{fig:mlcc_main}, 
which summarizes the overall GC--PLC--LC pipeline.
For clarity, Table~\ref{tab:notations} provides a summary of the main symbols used in the proposed framework.

The following sections detail the architecture across three tightly coupled stages:
\begin{itemize}
\item Section~\ref{subsec:compress} describes the global compressor and local compressor, which serve complementary roles in the distillation hierarchy—integrating holistic context and extracting localized essence, respectively.
\item Section~\ref{subsec:cross} introduces the progressive layered crossing module, a novel and efficient token interaction module that achieves expressive cross-token fusion while maintaining scalability and modularity.
\item Section~\ref{subsec:scale} discusses how the architecture can be extended along multiple scalability axes, and extends MLCC to a multi-channel variant termed \textbf{MC-MLCC}, offering a high ROI in terms of performance gain versus computational overhead.
\end{itemize}

\begin{table}
    \centering
    \captionsetup{skip=3pt}
    \caption{Notations}
    \begin{tabular}{c|c}
    \toprule
    Symbols & Meanings \\
    \midrule
        $N$ & number of original tokens \\
        $E$ & dimension of original tokens \\
        $H$ & number of global tokens (heads) \\
        $W$ & dimension of global tokens \\
        $K$ & number of hidden layers in dynamic MLP of PLC \\
        $e_i$ & output size of $i$-th layer in dynamic MLP of PLC \\
        $L$ & dimension of interwoven tokens \\
        $E'$ & dimension of refined tokens \\
        $S$ & number of channels \\
    \hline
        $\mathbf{X}$ & original tokens \\
        $\mathbf{M}$ & global tokens \\
        $\mathbf{C}$ & interwoven tokens \\
        $\mathbf{X}'$ & refined tokens  \\
        $\widetilde{\mathbf{X}}$ & multi-channel original tokens \\
        \bottomrule
    \end{tabular}
    \label{tab:notations}
\end{table}

\subsection{Compress: Multi-Level Compression Layers}
\label{subsec:compress}

The set of \emph{original tokens} $\mathbf{X} \in \mathbb{R}^{N \times E}$ consists of $N$ tokens, each represented as an embedding vector of dimension $E$, derived from the raw input feature fields.

To obtain a compact representation suitable for global reasoning, the \emph{global compressor} (GC) transforms $\mathbf{X}$ into a condensed set of \emph{global tokens} $\mathbf{M} \in \mathbb{R}^{H \times W}$:
\begin{equation}
    \mathbf{M}_{k,l}=\sum_{i=1}^{N} \sum_{j=1}^{E} \mathbf{X}_{i,j}  \mathbf{W}^{\mathrm{GC}}_{i,j,k,l} \quad (1\le k\le H, 1\le l \le W)  \label{equ:global_compress}
\end{equation}
where $\mathbf{W}^{\mathrm{GC}} \in \mathbb{R}^{N \times E \times H \times W}$ denotes trainable parameters that allow each global token to aggregate information across all original tokens.
Each global token thus encodes a holistic abstraction, acting as a condensed semantic summary of the entire input.
Conceptually, GC functions as a semantic projector, mapping scattered token-level information into a globally contextualized latent space that facilitates high-level interaction modeling.

Next, the original tokens and global tokens are fed into a dedicated feature interaction module to produce a new representation $\mathbf{C} \in \mathbb{R}^{N \times L}$, referred to as the \emph{interwoven tokens}, as defined in Equation~\ref{equ:mlcc_2}. Details of this interaction mechanism are described in Section~\ref{subsec:cross}.
 
Finally, a second compression step is applied via the \emph{local compressor} (LC), which projects the interwoven tokens into a refined representation $\mathbf{X}' \in \mathbb{R}^{N \times E'}$, where $E'$ denotes the output embedding dimension. The projection is parameterized by $\mathbf{W}^{\mathrm{LC}} \in \mathbb{R}^{N \times L \times E'}$:
\begin{equation}
    \mathbf{X}'_{k,j} =\sum_{i=1}^{L} \mathbf{C}_{k , i} \mathbf{W}^{\mathrm{LC}}_{k , i, j} \quad (1\le k\le N, 1\le j\le E') \label{equ:local_compress}.
\end{equation}
Unlike GC, LC operates in a token-wise manner, independently compressing each token representation, thereby preserving localized semantic details while removing redundant interaction patterns.

\paragraph{Information-Processing Perspective.}
Both GC and LC perform dimensionality reduction and signal refinement, but at different semantic levels.
GC performs global abstraction, compressing across tokens to construct a shared semantic space for collective reasoning, whereas LC performs local filtration, distilling token-specific interactions to preserve individual semantic identity while improving robustness.
Together, they form a hierarchical distillation pipeline from global contextualization to localized purification, enabling expressive yet computation-efficient token representations.

\subsection{Cross: Dynamic High-Order Token Interaction}
\label{subsec:cross}

Conventional cross operations, such as feature-wise inner products or element-wise multiplications, are inherently limited to capturing low-order interactions. 
Although stacking multiple such layers can, in principle, model higher-order relationships, this approach is often inefficient in practice, both computationally and in parameter usage, particularly when interaction patterns are sparse or highly non-linear.
To address these issues, we propose the PLC layer, which enables high-order token interactions in a single layer by generating and applying dynamic, multi-layer transformations conditioned on the input.
The detailed structure of PLC is illustrated in Figure~\ref{fig:mlcc_plc}.

After the global compressor, we obtain a set of global tokens $\mathbf{M} \in \mathbb{R}^{H \times W}$, 
where $H$ denotes the number of heads and $W$ is the token dimension. 
We define $W = \sum_{i=1}^K e_{i-1} e_i$, which corresponds to the parameterization size of a $K$-layer MLP with layer widths $\{e_i\}_{i=1}^K$ and input width $e_0 = E$.

For each head $h$, the corresponding global token $\mathbf{M}^h$ is reshaped into the parameters of a dynamic MLP:
\begin{equation}
    \left[\mathbf{D}^h_1,\dots,\mathbf{D}^h_K\right] = \mathbf{M}^h, 
    \quad \mathbf{D}^h_i \in \mathbb{R}^{e_{i-1} \times e_i}.
    \label{equ:plc_split}
\end{equation}
The dynamic MLP processes the original tokens $\mathbf{X}$ progressively:
\begin{equation}
    \mathbf{O}^h_k = \sigma\left(\mathbf{O}^h_{k-1} \mathbf{D}^h_k\right) \quad \left(1\le k\le K, \ \mathbf{O}^h_0=\mathbf{X}\right),
    \label{equ:plc_dmlp}
\end{equation}
where $\sigma(\cdot)$ denotes an optional activation function.  
The multi-level interaction outputs of head $h$ are then concatenated:
\begin{equation}
    \mathbf{O}^h = \text{Concat}\left(\mathbf{O}^h_1,\dots,\mathbf{O}^h_K\right).
    \label{equ:plc_concat1}
\end{equation}

Finally, the interwoven tokens are obtained by concatenating the original tokens with the outputs from all heads:
\begin{equation}
    \mathbf{C} = \text{Concat}\left(\mathbf{X}, \mathbf{O}^1, \dots, \mathbf{O}^H\right),
    \label{equ:plc_concat2}
\end{equation}
resulting in a token size of $L = E + H\sum_{i=1}^K e_i$.
This design jointly preserves low-order information from $\mathbf{X}$ and high-order interactions from the progressive transformations, without requiring deep stacking.

Equations~\ref{equ:plc_split}–\ref{equ:plc_concat2} can be summarized compactly as Equation~\ref{equ:mlcc_2}.

\paragraph{Attention-Inspired Perspective.}
PLC can also be interpreted through the lens of attention mechanisms.
The global tokens act as shared keys encoding global context, while the original tokens serve as queries.
Instead of computing scalar attention weights via dot-products, PLC replaces them with learned multi-layer transformation functions.
This formulation is analogous to a multi-query attention (MQA) setup, where each query–key interaction is expanded into a high-order nonlinear mapping.
PLC thus functions as a context-conditioned interaction mechanism that fuses global context with token-specific semantics to produce the interwoven tokens.

\subsection{Scale: Multi-Axis Scalability}
\label{subsec:scale}

As discussed in Section~\ref{subsec:cross}, conventional cross structures primarily increase model capacity through \emph{vertical stacking}, i.e., by deepening the network. 
Although deeper architectures can in principle capture higher-order interactions, they are often inefficient in practice: computational overhead grows rapidly, parameter usage becomes excessive, and empirical gains tend to saturate. 
More importantly, such designs lack principled mechanisms for \emph{horizontal expansion}, making it difficult to scale capacity without proportionally increasing depth.

In contrast, the proposed \emph{GC--PLC--LC} pipeline in MLCC is inherently amenable to horizontal scaling. 
Its modular structure allows model capacity to be expanded along multiple independent axes, enabling flexible and efficient scaling. 
We consider three complementary scaling strategies.

A straightforward approach is to increase the intrinsic capacity of the PLC module. 
Each PLC head instantiates a dynamic multi-layer perceptron (MLP), whose depth and width control the complexity of token interactions. 
By enlarging these dimensions, PLC can model richer high-order and non-linear cross patterns.

Another axis of expansion is the number of PLC heads. 
Each head corresponds to an independent dynamic MLP conditioned on global tokens. 
Increasing the head number $H$ enhances the diversity of learned interaction functions, thereby improving representational power without deep stacking.

Beyond modifying PLC itself, we further introduce multiple parallel \emph{channels}, as illustrated in Figure~\ref{fig:mlcc_multi}.
Specifically, the input embeddings are partitioned as
\begin{equation}
    \left[\mathbf{X}_1,\dots,\mathbf{X}_S\right] = \widetilde{\mathbf{X}},
\end{equation}
and each channel independently applies a GC--PLC pipeline:
\begin{equation}
    \mathbf{C}_i = \text{PLC}_i\left(\mathbf{X}_i;\text{GC}\left(\mathbf{X}_i\right)\right)\quad(1\le i \le S).
\end{equation}
Here, $S$ denotes the number of parallel channels, and each channel processes a disjoint partition of the token embeddings. 
Intuitively, a channel can be viewed as an independent interaction pathway that specializes in a subspace of the embedding space, enabling diverse relational structures to be modeled in parallel.

The resulting interwoven tokens are aggregated by concatenation,  
\begin{equation}
    \mathbf{C} = \text{Concat}(\mathbf{C}_1,\dots,\mathbf{C}_S), \label{equ:scale_concat}
\end{equation}
forming the combined representation\footnote{To avoid redundant concatenation of the original tokens across channels, the symbol $\mathbf{X}$ in Equation~\ref{equ:plc_concat2} can be shifted to Equation~\ref{equ:scale_concat}.}. 
The concatenated tokens are then refined by a single LC module, yielding the final representation $\mathbf{X}'$.

In this multi-channel setting, $\mathbf{X}_i$ and $\mathbf{C}_i$ maintain the same shapes as in the base GC and PLC modules, i.e., $\mathbb{R}^{N \times E}$ and $\mathbb{R}^{N \times L}$. 
The expanded tensors $\widetilde{\mathbf{X}}$, $\mathbf{C}$, and $\mathbf{W}_2$ thus take shapes $\mathbb{R}^{NS \times E}$, $\mathbb{R}^{NS \times L}$, and $\mathbb{R}^{NS \times L \times E'}$, respectively.

Taken together, these scaling strategies demonstrate that GC--PLC--LC supports orthogonal axes of expansion, including internal PLC capacity, head multiplicity, and parallel channels. 
Unlike traditional cross networks that rely solely on vertical depth, MLCC enables structured horizontal scaling, providing fine-grained control over the trade-off between expressiveness and efficiency.

We refer to this multi-channel extension as \textbf{M}ulti-\textbf{C}hannel MLCC (MC-MLCC), which is adopted as the strongest variant in all subsequent experiments.

\paragraph{Regulating Scalability via Adaptive Compression.}
Scaling along the above axes increases the dimensionality of token representations, which may exceed the capacity of downstream components. 
To regulate model size and computational cost, we define a \emph{compression ratio} $r = {E_{\mathrm{in}}}/{E_{\mathrm{out}}}$,
where $E_{\mathrm{in}}$ denotes the embedding length entering the interaction module and $E_{\mathrm{out}}$ denotes the refined embedding length after LC.
For single-channel MLCC, $E_{\mathrm{in}} = E$ and $E_{\mathrm{out}} = E'$, yielding $r = E/E'$. 
For multi-channel MC-MLCC, concatenation results in $E_{\mathrm{in}} = S \cdot E$, and thus $r = {S\cdot E}/{E'}$.

This definition makes $r$ an architecture-level control variable for comparing capacity--compression trade-offs across different scaling configurations. 
Increasing $r$ enforces stronger compression in LC, reducing parameter count and computation while preserving enriched high-order interactions produced by scaling.

\section{Experiments}
\label{sec:experiments}

In this section, we evaluate the proposed MLCC and its multi-channel variant MC-MLCC against representative CTR prediction baselines. 
We further conduct extensive ablation studies and scaling analyses to understand the effects of key architectural components and scaling axes.

\subsection{Experimental Setup}
\subsubsection{Models}
We compare MLCC and MC-MLCC with the following representative models:
\begin{itemize}
    \item \textbf{DNN} \cite{cheng2016wide}: the standard multilayer perceptron for DLRM.
    \item \textbf{DCNv2} \cite{wang2021dcn}: a model that captures explicit feature crosses via cross layers.
    \item \textbf{Wukong} \cite{zhang2024wukong}: a strong token-based baseline leveraging attention and compact embedding.
    \item \textbf{RankMixer} \cite{zhu2025rankmixer}: a recent mixer-style model based on token interactions.
\end{itemize}
Unless otherwise specified, all models share the same base embedding dimension $E$ to ensure fair comparison.

\subsubsection{Datasets}
\begin{table}[htbp]
\centering
\captionsetup{skip=3pt}
\caption{Statistics of public datasets used in our experiments.}
\label{tab:dataset_stats}
\begin{tabular}{lccc}
\toprule
\textbf{Dataset} & \textbf{\#Samples} & \textbf{\#Features} & \textbf{Positive Ratio} \\
\midrule
Criteo & 45M & 39 & 0.26 \\
Avazu & 40M & 22 & 0.17 \\
TaobaoAds & 26M & 17 & 0.05 \\
\bottomrule
\end{tabular}
\end{table}

We use four datasets:
\begin{itemize}
    \item \textbf{Criteo}\footnote{\url{https://www.kaggle.com/c/criteo-display-ad-challenge}}: a widely used public CTR benchmark dataset released by Criteo.
    \item \textbf{Avazu}\footnote{\url{https://www.kaggle.com/c/avazu-ctr-prediction}}: a large-scale CTR prediction dataset released by Avazu, consisting of user click records from mobile advertising.
    \item \textbf{TaobaoAds}\footnote{\url{https://tianchi.aliyun.com/dataset/56}}: an advertising dataset collected from the Taobao platform, including user-item interactions in display advertising scenarios.
    \item \textbf{Industrial Dataset}: a proprietary dataset from Bilibili in-feed advertising, containing anonymized user-item interaction logs.
\end{itemize}

Each public dataset is randomly split into training, validation, and test sets with a ratio of 90:1:9. 
Table~\ref{tab:dataset_stats} summarizes dataset statistics.

The industrial dataset provides a real-world benchmark with heterogeneous feature distributions and complex interaction patterns, making the overall evaluation more comprehensive in terms of both generalization and practical applicability of our proposed models.

\subsubsection{Evaluation Metrics}
We adopt a set of standard metrics to evaluate model performance:

\begin{itemize}
    \item \textbf{AUC} (Area under the ROC curve): This is our primary metric, reflecting the model's ranking quality and its ability to distinguish positive from negative samples.
    \item \textbf{LogLoss}: Binary cross-entropy loss, reporting prediction calibration. We provide LogLoss mainly for public datasets to facilitate comparisons with prior work.
    \item \textbf{\#Params}: Total number of trainable parameters, indicating model size and memory footprint.
    \item \textbf{FLOPs}: Floating-point operations per forward pass, measuring computational cost and serving as a proxy for inference efficiency.
\end{itemize}

In the following experiments, AUC is the main metric used to compare different models and configurations, while LogLoss, \#Params, and FLOPs provide complementary insights into model calibration, efficiency, and scalability.

\subsection{Main Results}

\begin{table*}[htbp]
\centering
\captionsetup{skip=3pt}
\caption{Performance comparison of different models on public datasets. 
Best results are in \textbf{bold}.}
\label{tab:main_results}
\begin{tabular}{lcccccc}
\toprule
\multirow{2}{*}{Model} & \multicolumn{2}{c}{Criteo} & \multicolumn{2}{c}{Avazu} &\multicolumn{2}{c}{TaobaoAds}  \\
\cmidrule(lr){2-3} \cmidrule(lr){4-5} \cmidrule(lr){6-7} 
 & AUC $\uparrow$ & LogLoss $\downarrow$ & AUC $\uparrow$ & LogLoss $\downarrow$ & AUC $\uparrow$ & LogLoss $\downarrow$   \\
\midrule
DNN & 0.7995 & 0.4554 & 0.7844 & 0.3781 & 0.6544 & 0.1949 \\
DCNv2 & 0.8012 & 0.4541 & 0.7879 & 0.3757 & 0.6556 & 0.1949 \\
Wukong & 0.8013 & 0.4594 & 0.7893 & 0.3787 & 0.6568 & 0.1952 \\
RankMixer & 0.8008 & 0.4625 & 0.7849 & 0.3818 & 0.6555 & 0.1949 \\
\midrule
\textbf{MLCC (ours)} & 0.8020 & \textbf{0.4517} & 0.7896 & 0.3747 & 0.6588 & 0.1943  \\
\textbf{MC-MLCC (ours)} & \textbf{0.8034} & 0.4526 & \textbf{0.7945} & \textbf{0.3720} & \textbf{0.6596} & \textbf{0.1942}  \\
\bottomrule
\end{tabular}
\end{table*}

\begin{table}[htbp]
\centering
\captionsetup{skip=3pt}
\caption{Performance comparison of different models on our industrial dataset. 
Best results are in \textbf{bold}.}
\label{tab:indsutry_results}
\begin{tabular}{lcrr}
\toprule
Model & AUC $\uparrow$ & \#Params (M) & GFLOPs \\
\midrule
DNN & 0.8275 & 30.0  & 123.1 \\
DCNv2 & 0.8304 & 96.9  & 397.2  \\
Wukong & \textbf{0.8334} & 411.8 & 1,738.3 \\
RankMixer & 0.8322 & 883.2  & 3,620.6 \\
\midrule
\textbf{MLCC (ours)} & 0.8333 & 65.2 & 268.3 \\
\textbf{MC-MLCC (ours)} & \textbf{0.8334} & 15.5 & 64.9 \\
\bottomrule
\end{tabular}
\end{table}

Tables~\ref{tab:main_results} and \ref{tab:indsutry_results} summarize the performance of all models on public and industrial datasets, respectively. 
Across all datasets, our proposed MLCC already outperforms conventional baselines, demonstrating the effectiveness of the structured cross-feature interaction mechanism. 
The multi-channel extension MC-MLCC further improves predictive performance while maintaining high computational efficiency.

On public datasets, MLCC consistently achieves higher AUC compared with DNN, DCNv2, Wukong, and RankMixer.  
For example, on Criteo, MLCC attains an AUC of 0.8020, surpassing all baselines, while reducing LogLoss to 0.4517.  
MC-MLCC further boosts AUC to 0.8034, confirming that multi-channel expansion enhances the model’s representation capacity.  
Similarly, on Avazu, MLCC achieves an AUC of 0.7896, already higher than all baselines, and MC-MLCC further improves to 0.7945, reducing LogLoss to 0.3720.  
On the sparser and more imbalanced TaobaoAds dataset, MLCC reaches an AUC of 0.6588, outperforming the strongest baseline Wukong (0.6568), and MC-MLCC increases AUC to 0.6596, while slightly lowering LogLoss to 0.1942.  
These results indicate that MLCC alone is highly effective, and MC-MLCC consistently provides additional gains across diverse public CTR benchmarks.

On the industrial dataset, MLCC already demonstrates strong performance, achieving an AUC of 0.8333, surpassing DNN, DCNv2 and RankMixer, while using fewer parameters and moderate FLOPs.  
MC-MLCC further slightly improves the AUC to 0.8334, matching the performance of the strongest baseline Wukong.  
Importantly, both MLCC and MC-MLCC achieve this comparable performance while drastically reducing model size and computational cost.
MLCC already reduces parameters and FLOPs by approximately \textbf{6$\times$} compared with Wukong.
MC-MLCC further improves efficiency, requiring over \textbf{26$\times$} fewer parameters and FLOPs than Wukong, demonstrating its strong efficiency for large-scale industrial deployment.

In summary, these results collectively demonstrate three key points:  
1) MLCC alone consistently outperforms traditional baselines across public and industrial datasets, confirming the value of structured cross-feature interactions;  
2) MC-MLCC further leverages multi-channel representations to achieve additional performance gains while drastically reducing computational cost and memory footprint;  
3) the proposed architectures provide a practical and efficient approach for large-scale CTR prediction in both public benchmarks and industrial settings.

\subsection{Ablation Studies}

The ablation studies on the industrial dataset aim to verify that the observed gains of MLCC are driven by the proposed compress--cross interaction mechanism rather than pure parameter scaling, to assess the role of Local Compressor (LC) and Progressive Layered Crossing (PLC) modules.

\subsubsection{Effect of the Local Compressor}

\begin{table}[htbp]
\centering
\captionsetup{skip=3pt}
\caption{Ablation study on the Local Compressor (LC) under a fixed base embedding size of $E=32$.}
\label{tab:ablation_lc}
\begin{tabular}{lrcrr}
\toprule
Variant & $r$ &AUC $\uparrow$  & \#Params (M) & GFLOPs \\
\midrule
MLCC (w/o LC)  & - & 0.8296 & 28.4 & 117.2 \\
MLCC ($E'=8$)  & 4 & 0.8301 & \textbf{9.8} & \textbf{40.7} \\
MLCC ($E'=16$)  & 2 & 0.8300 & 13.3 & 55.1\\
MLCC ($E'=32$)  & 1 & \textbf{0.8303} & 20.3 & 83.7\\
\bottomrule
\end{tabular}
\end{table}

This subsection fixes the base embedding dimension $E=32$, and adjusts the output dimension $E'$ of LC to control the compression ratio $r$, aiming to verify the dual value of LC in performance preservation and cost reduction.

Table~\ref{tab:ablation_lc} shows that introducing LC consistently improves model quality compared to disabling it.
Even under aggressive dimensionality reduction (e.g., $r=4$), MLCC maintains nearly the same AUC as the uncompressed setting.
This indicates that LC is more than a simple dimensionality reducer. 
It performs token-wise structured transformation, filtering redundant noise introduced during PLC interactions and enhancing the validity of local semantic information for high-order interactions.
Notably, shrinking $E'$ from 32 to 8 barely impacts accuracy, yet the parameter count and GFLOPs are reduced by over 2$\times$.

In other words, MLCC retains strong predictive power even with aggressive compression via LC, enabling highly efficient configurations that balance representational compactness and performance, which is a key advantage for industrial deployment under resource constraints.

\subsubsection{PLC vs. Inner Product}

\begin{table}[htbp]
\centering
\captionsetup{skip=3pt}
\caption{Comparison between inner product and our proposed PLC module under a fixed embedding size of $E=16$.}
\label{tab:ablation_dynamic}
\begin{tabular}{lcr}
\toprule
Variant & AUC $\uparrow$ & \#Params (M)  \\
\midrule
DNN & 0.8260 & 9.6\\
MLCC (inner product, $H=64$) & 0.8283 & 13.2\\
MLCC (PLC, $H=16$) & \textbf{0.8290} & 13.5  \\
\bottomrule
\end{tabular}
\end{table}

\begin{table}[htbp]
\centering
\captionsetup{skip=3pt}
\caption{Effect of varying the number of heads $H$ in the GC module under a fixed embedding size of $E=16$.}
\label{tab:scale_head}
\begin{tabular}{lcrr}
\toprule
$H$ & AUC $\uparrow$ & \#Params (M) & GFLOPs\\
\midrule
$1$ & 0.8278 & 9.9 & 40.8\\
$2$ & 0.8284 & 10.1 & 41.8\\
$4$ & 0.8285 & 10.6  & 43.9\\
$8$ & 0.8288 & 11.6 & 48.1\\
$16$ & 0.8290 & 13.5 & 56.6\\
$32$ & \textbf{0.8293} & 17.4 & 73.4\\
\bottomrule
\end{tabular}
\end{table}

This subsection fixes the embedding size $E=16$ and controls the parameter budget of MLCC variants to be nearly identical, focusing on comparing the interaction effectiveness between the proposed PLC module and the conventional static inner product.

Table~\ref{tab:ablation_dynamic} shows that even with comparable capacity, the PLC module outperforms the static inner product.
The inner-product variant improves over DNN by introducing explicit pairwise interactions, but the PLC module further raises AUC to 0.8290.
This advantage stems from the fundamental difference in interaction mechanisms: the static inner product can only model linear pairwise dependencies, while PLC generates context-aware dynamic MLP transformations for each token pair, enabling the capture of richer and more flexible high-order nonlinear interactions.
Notably, the PLC module achieves better performance with far fewer heads (16 vs. 64 for the inner-product variant), verifying its higher interaction efficiency, which is another key design goal of the compress--cross paradigm.

In summary, the PLC module offers a strictly stronger interaction formulation than fixed similarity measures, delivering better predictive power without requiring additional parameters.
This reinforces its role as the core interaction component in the proposed architecture, directly supporting the effectiveness of the ``dynamic cross'' mechanism.

\subsection{Scaling Laws}

\begin{table*}[htbp]
\centering
\captionsetup{skip=3pt}
\caption{Performance comparison under matched effective embedding dimension ($E \times S$), showing how increasing embedding size ($E$) or channel count ($S$) contributes to total representational capacity.}
\label{tab:scale_unified}
\begin{tabular}{c crr crr crr}
\toprule
\multirow{2}{*}{$E \times S$} 
& \multicolumn{3}{c}{DNN ($S=1$)} 
& \multicolumn{3}{c}{MLCC ($S=1$)} 
& \multicolumn{3}{c}{MC-MLCC ($E=16$)} \\
\cmidrule(lr){2-4} \cmidrule(lr){5-7} \cmidrule(lr){8-10}
 & AUC $\uparrow$ & \#Params (M) & GFLOPs
 & AUC $\uparrow$ & \#Params (M) & GFLOPs
 & AUC $\uparrow$ & \#Params (M) & GFLOPs \\
\midrule
16  & 0.8260 & 9.6 & 39.3 & 0.8278 & 9.9 & 40.8 & 0.8278 & 9.9 & 40.8 \\
32  & 0.8272 & 16.4 & 67.2 &  0.8293 & 10.6 & 43.7 & 0.8288 & 10.2 & 42.0 \\
48  & 0.8267 & 23.2 & 95.2 & 0.8304 & 11.7 & 48.5 & 0.8300 & 10.5 & 43.6 \\
64  & \textbf{0.8275} & 30.0 & 123.1 & 0.8310 &  13.3 & 54.8 & 0.8305 & 10.8  & 44.6 \\
96  & 0.8270 & 36.8 & 151.0 & 0.8319 & 17.7 & 73.0 & 0.8310 & 11.4 & 47.1 \\
128 & 0.8267 & 43.6 & 234.8 & 0.8324 & 23.8 & 98.1 & 0.8313 & 12.0 & 49.7 \\
160 & 0.8266 & 70.9 & 290.7 & 0.8329 & 31.6 & 130.1 & 0.8322 & 12.6 & 52.2\\
192 & -- & -- & -- & 0.8331 & 41.1 & 169.2 & 0.8328 & 13.2 & 54.8 \\
256 & -- & -- & -- & \textbf{0.8333} & 65.2 & 268.3 & \textbf{0.8331} & 14.3 & 59.9 \\
\bottomrule
\end{tabular}
\end{table*}

We further investigate how model performance scales with respect to head number ($H$), embedding dimension ($E$), and channel count ($S$). 
All experiments in this section are conducted on the industrial dataset. 
Fig.~\ref{fig:auc_flops} visualizes the scaling behavior of MC-MLCC along different architectural axes.

\subsubsection{Scaling with Head Number ($H$)}

This subsection explores how the number of heads $H$ affects model performance, with the base embedding size fixed at $E=16$.
Table~\ref{tab:scale_head} shows a clear upward trend: increasing $H$ broadens the interaction subspaces available to the model, allowing it to capture a richer set of localized patterns. 
The gain from $H=1$ to $H=8$ is noticeable, suggesting that even moderate head parallelism can significantly enhance the expressiveness of the cross-structure.

Beyond this point, performance continues to improve but at a slower pace.
The best AUC is reached at $H=32$, indicating that sufficiently diversified heads allow the model to disentangle complex feature relationships more effectively.
At the same time, the parameter count and FLOPs grow steadily with $H$, reflecting a natural trade-off between model capacity and computational cost.

Overall, scaling $H$ provides a predictable performance improvement, with AUC growing monotonically and computational cost increasing steadily, suitable for scenarios with moderate resource increments.

\subsubsection{Scaling with Embedding Dimension (E)}

\begin{figure}
    \centering
    \includegraphics[width=0.85\linewidth]{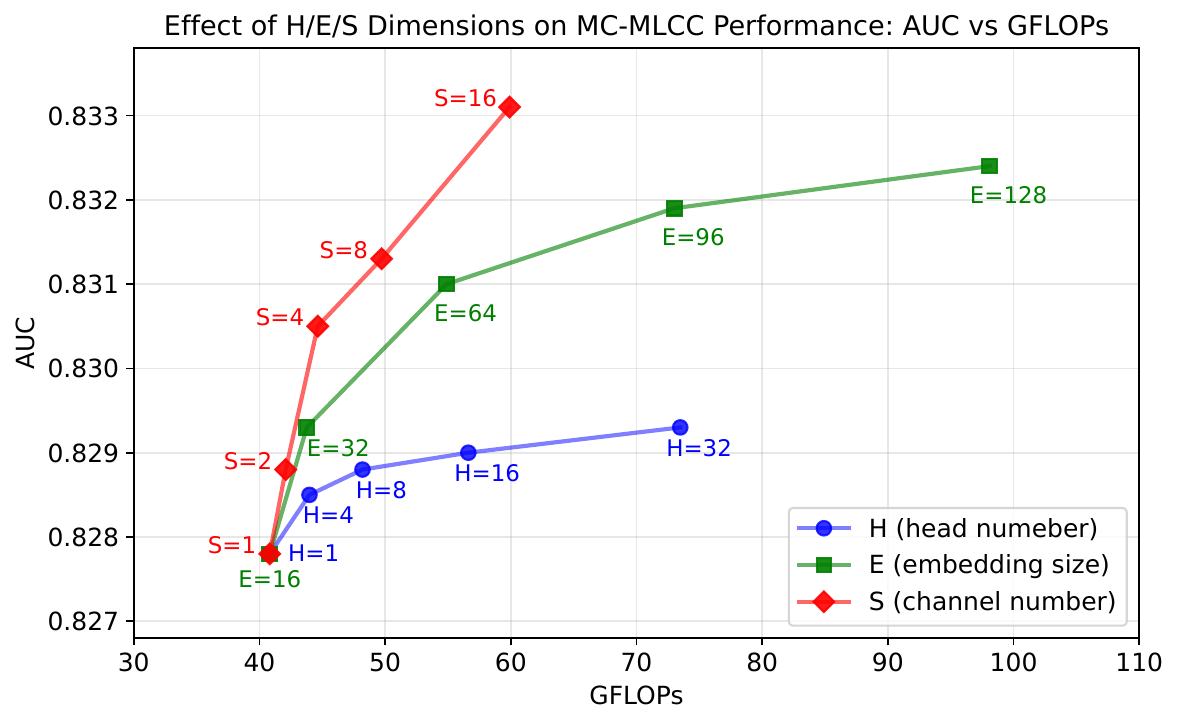}
    \captionsetup{skip=3pt}
    \caption{Effects of scaling different architectural axes in MC-MLCC on the industrial dataset. 
We report AUC versus GFLOPs when scaling (1) the number of heads $H$, (2) the embedding dimension $E$, and (3) the number of channels $S$.}
    \label{fig:auc_flops}
\end{figure}

Table~\ref{tab:scale_unified} summarizes how model performance varies with the embedding dimension $E$.
The comparison focuses on the \textbf{DNN ($S=1$)} and \textbf{MLCC ($S=1$)} columns, both of which scale capacity solely by increasing $E$.

For the DNN baseline, performance improves as $E$ grows from 16 to 64, but then plateaus and begins to fluctuate, suggesting that simply enlarging embeddings provides limited additional expressive power and may even introduce redundancy.  
This behavior is consistent with observations in prior recommender system studies, which report that high-dimensional embeddings often suffer from embedding collapse \cite{chen2024towards,guo2024embedding,peng2025balancing}.

In contrast, MLCC shows a steadier and more consistent upward trend as $E$ increases, achieving its best performance at $E=256$.
Because MLCC explicitly structures cross-feature interactions, it is able to capitalize on the additional representational capacity without overfitting or saturating as quickly as the DNN.

This trend is also reflected in Fig.~\ref{fig:auc_flops}, where MLCC maintains a more favorable AUC growth curve than DNN as the effective dimension increases, despite a comparable rise in computational cost.
Together, these results indicate that the benefit of larger embeddings critically depends on whether the model can translate raw capacity into structured and meaningful feature interactions.

More broadly, this comparison indicates that while higher embedding dimensions can help, their utility strongly depends on the model architecture’s ability to extract meaningful interactions from them.
MLCC demonstrates that structured interaction mechanisms can translate increased embedding size into more effective learning.

\subsubsection{Scaling with Channel Number ($S$)}

The comparison here focuses on \textbf{MLCC ($S=1$)} versus \textbf{MC-MLCC ($E=16$)} in Table~\ref{tab:scale_unified}, with both operating under matched effective dimensions ($E\times S$) but scaling capacity in fundamentally different ways: MLCC enlarges the embedding size $E$, while MC-MLCC expands parallel channels $S$.

This alternative scaling pathway reveals a distinct advantage of the multi-channel design.
As $S$ increases, MC-MLCC decomposes the representation space into multiple complementary subspaces, allowing the model to exploit greater effective dimensionality without the optimization difficulty or parameter inflation associated with large single embeddings. 
Notably, MC-MLCC with $E=16$ and $S=16$ (effective dimension 256) matches the performance of MLCC at $E=192$ while using substantially fewer parameters.
When further increasing the embedding size to 256, MLCC achieves a slightly higher AUC, but at the cost of a dramatic increase in model size and computation. 
In contrast, MC-MLCC at the same effective dimension attains comparable performance with over 4$\times$ fewer parameters and FLOPs, highlighting the superior parameter efficiency of channel-based scaling.
This illustrates how channel-based expansion can serve as a more efficient route to increase expressiveness, particularly when large embedding vectors become harder to train or yield diminishing returns.
Combined with Fig.~\ref{fig:auc_flops}, channel-scaled MC-MLCC not only achieves comparable AUC to embedding-scaled MLCC but also maintains consistently lower GFLOPs, indicating a more favorable scalability trajectory.

Taken together, the two scaling analyses suggest a coherent picture: while increasing embedding size can improve performance, distributing capacity across multiple channels offers a more robust and parameter-efficient alternative, especially at larger effective dimensions.

\subsubsection{Discussion on scaling-up ROI}

\begin{figure}
    \centering
    \includegraphics[width=0.85\linewidth]{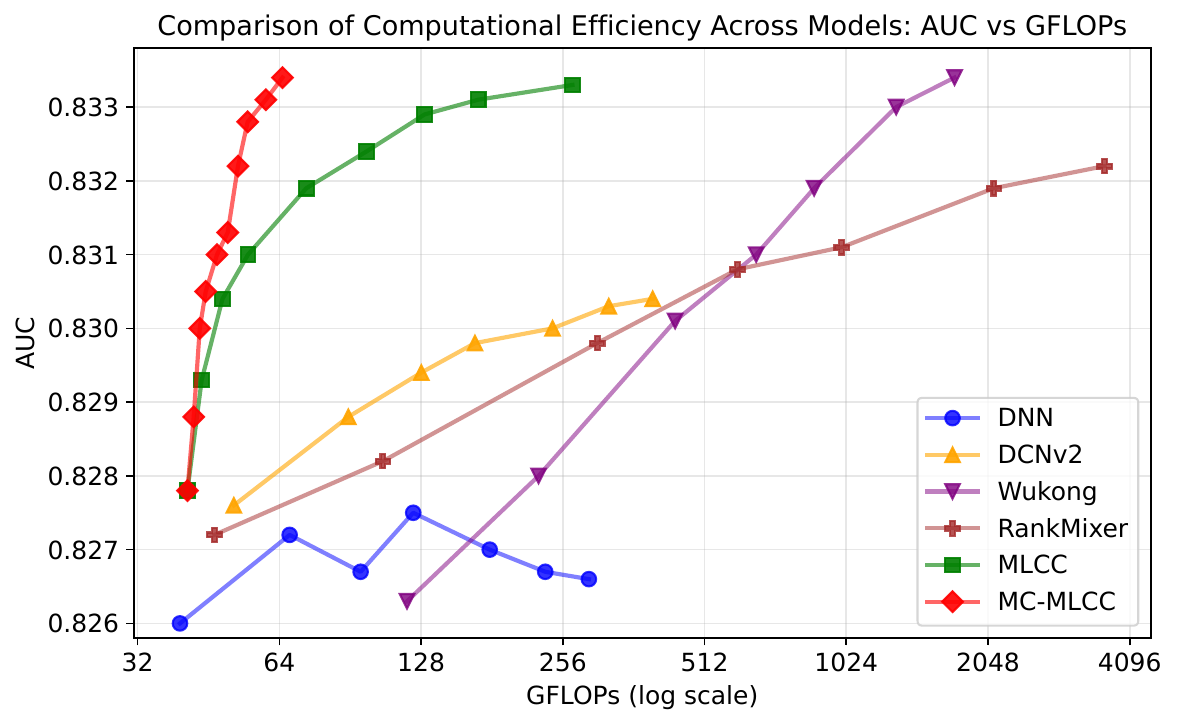}
    \captionsetup{skip=3pt}
    \caption{Computational efficiency comparison across models on the industrial dataset.}
    \label{fig:auc_flops_v2}
\end{figure}

Fig.~\ref{fig:auc_flops_v2} intuitively reflects the accuracy-efficiency trade-off advantage of our proposed architectures compared to mainstream models, which is the core basis for analyzing scaling-up ROI.
Under the same GFLOPs level, both MLCC and MC-MLCC achieve higher AUC than all other models.
When AUC is comparable, MC-MLCC requires drastically fewer GFLOPs than baseline models.
This efficiency advantage stems from the fundamental difference in scaling strategies: conventional models rely on embedding inflation or deep stacking, which leads to rapid growth in parameters and computation without proportional performance gains.
In contrast, our channel-based horizontal scaling and structured compress--cross mechanism enable more efficient conversion of computational resources into predictive performance.

The experiments on head number $H$, embedding dimension $E$, and channel count $S$ further confirm that MLCC and MC-MLCC exhibit superior scaling-up ROI compared to conventional models.
Performance improves monotonically as each scaling axis is increased, and the AUC gain per additional parameter is substantially higher.
Among all scaling strategies, channel expansion in MC-MLCC stands out with the best ROI.
It maintains a small embedding size per channel and distributes representation across multiple parallel channels, achieving comparable or even superior performance to embedding-scaled models while reducing parameter count and computational cost drastically.

This approach mitigates the optimization difficulty and memory overhead associated with high-dimensional embeddings, providing a more flexible and practical route for industrial deployment.
The predictable and favorable scaling behavior of our proposed architectures allows practitioners to select appropriate configurations based on computational budget, target ROI, and deployment constraints, ensuring both efficiency and strong predictive performance in real-world recommendation tasks.

\subsection{Online Experiments}
We deployed three iterative versions of our proposed framework within Bilibili advertising system, starting from its e-commerce vertical and progressively evolving from a compressed baseline to a fully scaled multi-channel architecture. 
Across these iterations, the framework delivered a cumulative \textbf{+32\%} improvement in ADVV, with inference latency maintained within production constraints.

Beyond the e-commerce vertical, MLCC variants have since been widely adopted across ranking models on Bilibili advertising platform. 
Cumulatively across these deployments, MLCC has resulted in over 9\% ADVV improvement at the platform level and has become a backbone interaction architecture in Bilibili commercial recommender systems. Detailed iteration trajectories and ablation studies for the e-commerce vertical are provided in Appendix~\ref{app:online}.

\section{Conclusion}

Balancing feature interaction capacity, computational efficiency, and scalability remains a core challenge in deep learning recommendation models (DLRMs). To address this, we propose a unified \emph{compress--cross--scale} paradigm with a novel architecture \textbf{MLCC}, and its multi-channel extension \textbf{MC-MLCC}, for efficient and scalable feature interaction modeling.

MLCC integrates hierarchical compression with dynamic cross modeling, efficiently capturing high-order feature dependencies without excessive overhead.
It outperforms strong DLRM baselines by 0.07\%–0.20\% AUC while reducing parameters and FLOPs by up to 6$\times$, validating the value of structured interaction mechanisms.
MC-MLCC introduces a multi-channel horizontal scaling pathway beyond embedding inflation.
By distributing interaction modeling across parallel subspaces, it matches top baseline performance with over \textbf{26$\times$} fewer parameters and FLOPs, and outperforms computation-matched DLRMs in AUC by up to \textbf{0.52\%}.
Scaling analyses confirm channel-based scaling delivers better ROI than traditional embedding expansion.

This work highlights that well-designed interaction architectures are key to efficient DLRM scaling.
The compress--cross--scale paradigm provides a principled foundation for balancing performance, efficiency and scalability, offering insights for industrial-grade recommender system research.

\appendix

\section{Online Iterations: A Progressive Scaling Journey}
\label{app:online}

\begin{table}[htbp]
\centering
\captionsetup{skip=3pt}
\caption{Three-Stage Online Evolution: Architectural Improvements Enable Progressive Embedding Scaling.}
\label{tab:online_evolution}
\begin{tabular}{lcc}
\toprule
\textbf{Model} & \textbf{ADVV Lift} & \textbf{Offline AUC Lift} \\
\midrule
\textbf{MLCC (inner product)}  & +9.03\% & +0.4\% \\
\textbf{MLCC (PLC)}  & +11.02\%  & +0.5\% \\
\textbf{MC-MLCC}  & +11.97\%  & +0.6\% \\
\bottomrule
\end{tabular}
\end{table}


This appendix documents the iterative online deployment of our framework on Bilibili e-commerce advertising platform.
The three-stage evolution from compression to enhanced interaction modeling and finally to horizontal scaling demonstrates how architectural improvements sequentially enable embedding expansions, validating our \emph{Compress, Cross and Scale} paradigm in production.
Each architectural improvement not only delivered immediate performance gains but also created computational and representational conditions for the next stage of scaling.

Table~\ref{tab:online_evolution} captures the three iterations, highlighting the symbiotic relationship between structural improvements and embedding scaling:
\begin{itemize}
    \item \textbf{MLCC (inner product)} replaced quadratic cross-feature operations with hierarchical compression (GC+LC). With a conservative embedding size of 8, it achieved +9.03\% ADVV lift while ensuring production feasibility.
    \item \textbf{MLCC (PLC)} replaced the inner product with PLC, i.e., a dynamic MLP, significantly enhancing interaction capacity. Crucially, this architectural improvement \textit{increased computational efficiency} relative to the expressiveness gained, creating headroom for a 4$\times$ embedding expansion (8$\to$32). The resulting +11.02\% ADVV lift validated that enhanced cross modeling and increased embedding dimensions work synergistically.
    \item \textbf{MC-MLCC} introduced multi-channel parallelism, enabling independent processing of diverse interaction subspaces.
This horizontal scaling strategy allowed another 4$\times$ embedding expansion (32$\to$128) with minimal computational overhead, delivering a further +11.97\% ADVV lift.
\end{itemize}

The cumulative outcome of +32\% ADVV and +1.5\% AUC improvement validates our framework's ability to evolve from computational efficiency to enhanced expressiveness through principled architectural innovations.
This trajectory exemplifies how a well-structured model can progressively scale in production while maintaining deployability.

\bibliographystyle{ACM-Reference-Format}
\bibliography{main}

@inproceedings{wang2017deep,
  title={Deep \& cross network for ad click predictions},
  author={Wang, Ruoxi and Fu, Bin and Fu, Gang and Wang, Mingliang},
  booktitle={Proceedings of the ADKDD'17},
  pages={1--7},
  year={2017}
}

@incollection{wang2021dcn,
  title={Dcn v2: Improved deep \& cross network and practical lessons for web-scale learning to rank systems},
  author={Wang, Ruoxi and Shivanna, Rakesh and Cheng, Derek and Jain, Sagar and Lin, Dong and Hong, Lichan and Chi, Ed},
  booktitle={Proceedings of the web conference 2021},
  pages={1785--1797},
  year={2021}
}

@inproceedings{qu2016product,
  title={Product-based neural networks for user response prediction},
  author={Qu, Yanru and Cai, Han and Ren, Kan and Zhang, Weinan and Yu, Yong and Wen, Ying and Wang, Jun},
  booktitle={2016 IEEE 16th international conference on data mining (ICDM)},
  pages={1149--1154},
  year={2016},
  organization={IEEE}
}

@inproceedings{lian2018xdeepfm,
  title={xdeepfm: Combining explicit and implicit feature interactions for recommender systems},
  author={Lian, Jianxun and Zhou, Xiaohuan and Zhang, Fuzheng and Chen, Zhongxia and Xie, Xing and Sun, Guangzhong},
  booktitle={Proceedings of the 24th ACM SIGKDD international conference on knowledge discovery \& data mining},
  pages={1754--1763},
  year={2018}
}

@inproceedings{song2019autoint,
  title={Autoint: Automatic feature interaction learning via self-attentive neural networks},
  author={Song, Weiping and Shi, Chence and Xiao, Zhiping and Duan, Zhijian and Xu, Yewen and Zhang, Ming and Tang, Jian},
  booktitle={Proceedings of the 28th ACM international conference on information and knowledge management},
  pages={1161--1170},
  year={2019}
}

@article{bian2020can,
  title={Can: Feature co-action for click-through rate prediction},
  author={Bian, Weijie and Wu, Kailun and Ren, Lejian and Pi, Qi and Zhang, Yujing and Xiao, Can and Sheng, Xiang-Rong and Zhu, Yong-Nan and Chan, Zhangming and Mou, Na and others},
  journal={arXiv preprint arXiv:2011.05625},
  year={2020}
}

@inproceedings{zhang2024wukong,
  title={Wukong: towards a scaling law for large-scale recommendation},
  author={Zhang, Buyun and Luo, Liang and Chen, Yuxin and Nie, Jade and Liu, Xi and Li, Shen and Zhao, Yanli and Hao, Yuchen and Yao, Yantao and Wen, Ellie Dingqiao and others},
  booktitle={Proceedings of the 41st International Conference on Machine Learning},
  pages={59421--59434},
  year={2024}
}

@inproceedings{cheng2016wide,
  title={Wide \& deep learning for recommender systems},
  author={Cheng, Heng-Tze and Koc, Levent and Harmsen, Jeremiah and Shaked, Tal and Chandra, Tushar and Aradhye, Hrishi and Anderson, Glen and Corrado, Greg and Chai, Wei and Ispir, Mustafa and others},
  booktitle={Proceedings of the 1st workshop on deep learning for recommender systems},
  pages={7--10},
  year={2016}
}

@inproceedings{guo2017deepfm,
  title={DeepFM: A Factorization-Machine based Neural Network for CTR Prediction},
  author={Guo, Huifeng and Ruiming, TANG and Ye, Yunming and Li, Zhenguo and He, Xiuqiang},
  booktitle={Proceedings of the Twenty-Sixth International Joint Conference on Artificial Intelligence},
  year={2017},
  organization={International Joint Conferences on Artificial Intelligence Organization}
}

@article{vaswani2017attention,
  title={Attention is all you need},
  author={Vaswani, Ashish and Shazeer, Noam and Parmar, Niki and Uszkoreit, Jakob and Jones, Llion and Gomez, Aidan N and Kaiser, {\L}ukasz and Polosukhin, Illia},
  journal={Advances in neural information processing systems},
  volume={30},
  year={2017}
}

@inproceedings{zhu2025rankmixer,
  title={Rankmixer: Scaling up ranking models in industrial recommenders},
  author={Zhu, Jie and Fan, Zhifang and Zhu, Xiaoxie and Jiang, Yuchen and Wang, Hangyu and Han, Xintian and Ding, Haoran and Wang, Xinmin and Zhao, Wenlin and Gong, Zhen and others},
  booktitle={Proceedings of the 34th ACM International Conference on Information and Knowledge Management},
  pages={6309--6316},
  year={2025}
}

@inproceedings{zhai2024actions,
  title={Actions speak louder than words: trillion-parameter sequential transducers for generative recommendations},
  author={Zhai, Jiaqi and Liao, Lucy and Liu, Xing and Wang, Yueming and Li, Rui and Cao, Xuan and Gao, Leon and Gong, Zhaojie and Gu, Fangda and He, Jiayuan and others},
  booktitle={Proceedings of the 41st International Conference on Machine Learning},
  pages={58484--58509},
  year={2024}
}

@inproceedings{han2025mtgr,
  title={Mtgr: Industrial-scale generative recommendation framework in meituan},
  author={Han, Ruidong and Yin, Bin and Chen, Shangyu and Jiang, He and Jiang, Fei and Li, Xiang and Ma, Chi and Huang, Mincong and Li, Xiaoguang and Jing, Chunzhen and others},
  booktitle={Proceedings of the 34th ACM International Conference on Information and Knowledge Management},
  pages={5731--5738},
  year={2025}
}

@article{chen2024hllm,
  title={Hllm: Enhancing sequential recommendations via hierarchical large language models for item and user modeling},
  author={Chen, Junyi and Chi, Lu and Peng, Bingyue and Yuan, Zehuan},
  journal={arXiv preprint arXiv:2409.12740},
  year={2024}
}

@article{rajput2023recommender,
  title={Recommender systems with generative retrieval},
  author={Rajput, Shashank and Mehta, Nikhil and Singh, Anima and Hulikal Keshavan, Raghunandan and Vu, Trung and Heldt, Lukasz and Hong, Lichan and Tay, Yi and Tran, Vinh and Samost, Jonah and others},
  journal={Advances in Neural Information Processing Systems},
  volume={36},
  pages={10299--10315},
  year={2023}
}

@article{deng2025onerec,
  title={Onerec: Unifying retrieve and rank with generative recommender and iterative preference alignment},
  author={Deng, Jiaxin and Wang, Shiyao and Cai, Kuo and Ren, Lejian and Hu, Qigen and Ding, Weifeng and Luo, Qiang and Zhou, Guorui},
  journal={arXiv preprint arXiv:2502.18965},
  year={2025}
}

@inproceedings{guo2024embedding,
  title={On the embedding collapse when scaling up recommendation models},
  author={Guo, Xingzhuo and Pan, Junwei and Wang, Ximei and Chen, Baixu and Jiang, Jie and Long, Mingsheng},
  booktitle={Proceedings of the 41st International Conference on Machine Learning},
  pages={16891--16909},
  year={2024}
}

@inproceedings{chen2024towards,
  title={Towards mitigating dimensional collapse of representations in collaborative filtering},
  author={Chen, Huiyuan and Lai, Vivian and Jin, Hongye and Jiang, Zhimeng and Das, Mahashweta and Hu, Xia},
  booktitle={Proceedings of the 17th ACM international conference on web search and data mining},
  pages={106--115},
  year={2024}
}

@article{peng2025balancing,
  title={Balancing Embedding Spectrum for Recommendation},
  author={Peng, Shaowen and Sugiyama, Kazunari and Liu, Xin and Mine, Tsunenori},
  journal={ACM Transactions on Recommender Systems},
  volume={3},
  number={4},
  pages={1--25},
  year={2025},
  publisher={ACM New York, NY}
}

@article{he2025understanding,
  title={Understanding Embedding Scaling in Collaborative Filtering},
  author={He, Yicheng and Kaiyu, Zhou and Bai, Haoyue and Zhu, Fengbin and Yang, Yonghui},
  journal={arXiv preprint arXiv:2509.15709},
  year={2025}
}

@article{ardalani2022understanding,
  title={Understanding scaling laws for recommendation models},
  author={Ardalani, Newsha and Wu, Carole-Jean and Chen, Zeliang and Bhushanam, Bhargav and Aziz, Adnan},
  journal={arXiv preprint arXiv:2208.08489},
  year={2022}
}

@inproceedings{zhang2024scaling,
  title={Scaling law of large sequential recommendation models},
  author={Zhang, Gaowei and Hou, Yupeng and Lu, Hongyu and Chen, Yu and Zhao, Wayne Xin and Wen, Ji-Rong},
  booktitle={Proceedings of the 18th ACM Conference on Recommender Systems},
  pages={444--453},
  year={2024}
}

@inproceedings{xu2024enhancing,
  title={Enhancing Performance and Scalability of Large-Scale Recommendation Systems with Jagged Flash Attention},
  author={Xu, Rengan and Yang, Junjie and Xu, Yifan and Li, Hong and Liu, Xing and Shankar, Devashish and Zhang, Haoci and Liu, Meng and Li, Boyang and Hu, Yuxi and others},
  booktitle={Proceedings of the 18th ACM Conference on Recommender Systems},
  pages={778--780},
  year={2024}
}

@inproceedings{ye2025fuxi,
  title={FuXi-$\alpha$: Scaling Recommendation Model with Feature Interaction Enhanced Transformer},
  author={Ye, Yufei and Guo, Wei and Chin, Jin Yao and Wang, Hao and Zhu, Hong and Lin, Xi and Ye, Yuyang and Liu, Yong and Tang, Ruiming and Lian, Defu and others},
  booktitle={Companion Proceedings of the ACM on Web Conference 2025},
  pages={557--566},
  year={2025}
}

@article{xu2025addressing,
  title={Addressing Information Loss and Interaction Collapse: A Dual Enhanced Attention Framework for Feature Interaction},
  author={Xu, Yi and Lu, Zhiyuan and Li, Xiaochen and Hu, Jinxin and Wen, Hong and Chen, Zulong and Zhang, Yu and Zhang, Jing},
  journal={arXiv preprint arXiv:2503.11233},
  year={2025}
}

@inproceedings{loveland2025understanding,
  title={Understanding and Scaling Collaborative Filtering Optimization from the Perspective of Matrix Rank},
  author={Loveland, Donald and Wu, Xinyi and Zhao, Tong and Koutra, Danai and Shah, Neil and Ju, Mingxuan},
  booktitle={Proceedings of the ACM on Web Conference 2025},
  pages={436--449},
  year={2025}
}

\end{document}